\documentclass[conference]{IEEEtran}
\IEEEoverridecommandlockouts
\usepackage{amsmath,amssymb,amsfonts}
\usepackage{algorithmic}
\usepackage{graphicx}
\usepackage{textcomp}
\usepackage[utf8]{inputenc}
\usepackage{graphicx}
\usepackage{hyperref}
\usepackage{amsmath}
\usepackage{amsfonts}
\usepackage{amssymb}
\begin{document}
\title{Security Analysis of Big Data on Internet of Things} 

\author{\IEEEauthorblockN{1\textsuperscript{st} Saeed Banaeian Far}
\IEEEauthorblockA{\textit{Department of Electrical Engineering,}\\ \textit{Yadegar -e- Imam Khomeini (rah), shahr-e-rey Branch} \\
\textit{Islamic Azad University}\\
Tehran, Iran \\
Saeed.banaeian@srbiau.ac.ir}
 \and 
\IEEEauthorblockN{2\textsuperscript{nd} Azadeh Imani Rad}
\IEEEauthorblockA{\textit{Department of Electrical Engineering,}\\ \textit{Yadegar -e- Imam Khomeini (rah), shahr-e-rey Branch} \\
\textit{Islamic Azad University}\\
Tehran, Iran \\
Azadeh\_{Imany}@yahoo.com}}

\maketitle

\begin{abstract}
The volume of data exchange in this network is ascending, by expansing of internet of things and its increasing familiarity in recent years. By increasing of requests for joining to the network and taking advantage of its services, necessity to maintain the privacy and security is felt more than ever and the whole thing has changed into a challenge. Keeping the security of the users in small networks seems to be simpler and the threats more predictable. As the users of the network which are sensors, the volume of data increases as well.  As a result the routers of  the network and network servers has become a seriously challenging task to control and maintain the security for this volume of data. We analyze these challenges in this paper. This paper aims to maintain the security of the users and present procedures to dominate the problems and express the strategies to overcome them. Some methods such as using both encryption function and secure protocol are proposed. Other method to overcome over big data challenges is smart design of communication protocol. Finally, we point out some future challenges of big-data on the internet of thing network.
\end{abstract}
\section{Introduction}
Not only the number of individuals connected to the internet increases day by day, but also the same thing happens to the things connected to the internet to exchange data and interact with each other. In modern era, smart devices and things could transmit and receive information as well as interact with each other. This interaction has reached a level that a major volume of data exchange through things is related to Internet of Things (IoT). It is forecasted that the number of smart things connected to internet in the year $2020$ is going to increase $44$ times comparing to $2009$ \cite{1}. The sensors could be enumerated as the dominant things connected to internet exchanging data, considered as one of the greatest resources of data \cite{2}. IoT shall be considered as a smart network consisting of things interacting with each other \cite{3}.\\ 
Things in the network produce and exchange data having a great volume, velocity, and variety. A term titled \textit{big data} (BD) has been coined for the data embodying the above mentioned characteristics. BD has no clear boundary and is usually classified based on its specifications \cite{4,5}.
\begin{itemize}
\item \textbf{Related work:} In $2015$, the importance of BD was investigated in the industries and got to the conclusion that complexity is the most important challenge of BD \cite{6}. Another survey carried out in the same year, considered the sensors and radio frequency identification systems (RFID) as the most important resources for the generation of BD \cite{7}. In $2016$, Jacobik investigated some security challenges and privacy settings of the BD \cite{8}. In the same year Ye \textit{et al}., investigated and presented a classification of attacks in various sectors of BD. In $2017$. Khan presented the utilization of the fourth generation of industrial revolution of BD \cite{9}. In this scheme, the IoT which is one of the generating factors of BD plays an important role which has been shortened as Industrial Internet of Thing (IIoT). In the same year, Sivaraja has expounded some challenges related to data, processing, data management, and afterwards has investigated some types of BD \cite{10}.
\item \textbf{Our contribution:} To overcome on BD challenges, we suggest several methods, as follows:
\begin{itemize}
\item \textit{Using encryption functions}
\item \textit{Network security}
\item \textit{Protecting programs}
\item \textit{Using secure protocol}
\item \textit{Using from lightweight encryption functions} 
\item \textit{Smart design of communication protocols} 
\end{itemize}
In the section $5$, we discuss about the mentioned methods.
\item \textbf{Paper organization:} This paper, the pre-requisites of work security and communication in IoT and also BD would be expounded after presenting some required definitions in the section $2$ and further scientific familiarization with the fundamentals of IoT principles as well as work security and communication in the network of IoT and also, the BD. Afterwards, in the section $4$ of the article the existing challenges in BD communication in IoT would be expounded. In the section $5$ of the article, some methods to overcome the challenges would be expressed from the viewpoint of network security. Finally, after the challenges facing to BD communication in IoT would be expressed.
\end{itemize}
\section{Preliminaries and Definitions}
Before entering detailed discussion of IoT more familiarization is needed. Concepts such as the network we discuss, instruments utilized or cases which faced with. In continuation, a brief description of IoT is needed to be explained.
\subsection{Internet of Things}
IoT is said to all things which are able to transfer or receive information and could interact with each other. IoT could be termed as an expanded and pervasive in all fields. This network is capable of gathering, processing, transmitted data analyzing of all network elements. IoT predicts that in future the world would be controlled by physical things while are connected to each other through a single infra-structure3 \cite{3}. In other words, IoT is a giant network in which a huge mass of things, sensors, and smart devices are interactive with each other \cite{11}. Most of this network is made up by sensors. But, there are other entities like $RFID$, smart card, smart phone, and the computers which are share the required information.
\subsection{Big Data}
No clear and transparent definition could be provided from BD \cite{5}. But as evident from the term \textit{big data} which refers to data containing high volumes of data, the most important point is the volume of data \cite{13}. Many definitions have been presented which are dependent upon the objective and research area. At the moment, $2$ billion individuals are connected to internet and $5$ billion use cellular phones. It is expected that the number of devices connected to the internet exceeds $50$ billion devices by $2020$. It is predicted that the number of messages transmitted by this year ($2020$) exceeds $44$ times the number of messages transmitted by $2009$, the imagination of this volume of change is thoroughly difficult.  The rate of BD is not clear, but some BD producers are You-tube, Facebook and Twitters \cite{1}. Table \ref{source} shows a number of other BD producers.\\
Nowadays, wireless sensor networks could be considered as BD producers as well \cite{2,7}. Predict that IoT network has the largest shared document, in future.
%%%%%%%%%%%%%%%%%%%%%%%%%%%%%%%%%%%%%%%%%%
\begin{table}
\caption{Some big data producers \cite{1}}
\centering
\resizebox{\linewidth}{!}{
\begin{tabular}{|c|l|}
\hline
\textbf{Source} & \textbf{Data Type and Size}  \\
\hline
YouTube & $1.$ Upload over $100$ $video/min$\\
  & $2.$ $1$ billion access $/mon$ \\
  & $3.$ Download $6$ billion $hour.video/mon$ \\
\hline  
Facebook & $1.$ $35$ billion likes $/min$\\
  & $2.$ Upload $100$ TB of data $/day$\\
  & $3.$ $1.4$ billion of users\\
\hline  
Twitter & $1.$ $645$ billion users\\
  & $2.$ $175$ billion twits $/day$\\
\hline
Google$+$ & $1.$ $1$ billion users \\
\hline
Google & $1.$  $2$ million search $/min$\\
  & $2.$ Process $25$ PB of data $/day$ \\
\hline
Apple & $1.$ $47000$ application download $/min$ \\ 
\hline
\end{tabular}
}
\label{source}
\end{table}
%%%%%%%%%%%%%%%%%%%%%%%%%%%%%%%%%%
\subsection{Cloud Server and Cloud Computing}
The information systems based on IoT are stored and processed in servers which have ultra-high strength and power. Cloud servers collect their information from sensors, $RFID$s and other smart devices and store them on their memory \cite{12}. Cloud computations bore growth from industry and the powerful servers of companies like Google and Amazon supported these servers \cite{11}. The need for cloud servers in IoT network is strongly felt, since this network involves light devices and they can not process and store BD.
\subsection{Big Data Application on Internet of Things}
The role of BD in IoT or the role of IoT in BD$?$ In order to reply this question it could somehow be told that the two phrases are interdependent together. The clear sample of BD could be observed through BD in IoT network. Also, several years ago, social networks such as Facebook were considered as giant producers of BD, and the social media have still a noticeable role, but predictions show that in a near future, IoT gains the highest share of exchanged information. As mentioned above, future attitude toward IoT focuses on large-scale smart cities \cite{14}. For example, in $2015$, Sun presented a smart city in four parts: $1)$Viability, $2)$protection, $3)$Rejuvenation, and $4)$Stability. In this scheme, the lowest level of architecture appropriate for rendering services is transmission of information through sensors.  After investigating information in the scheme, the time for client services is announced.\\
From the other applications is the transfer of industrial information. The fourth industrial revolution in the world in $2011$ was the apex of IoT usage.  Before that it was through robots, computers and chips. It was here that IoT entered to assist industry. The sensors which have the highest share in IoT started to increase \cite{15}. Along with this increase, the volume of exchanged controlled information ascended as well.\\
Another connection between IoT and BD, gathering and environmental information, GIS and astronomy through wireless sensors of IoT. With the increase of IoT, this information. Ascends as well \cite{16}.\\
Philip Chen has expounded another type of BD dependence to IoT \cite{13}. Navigation, Social media, financial information, Health information,  astronomy information, and smart transportation are among industries which produce lots of information.
\section{Requirements }
With demand increase in social media and information. Exchange in these networks, and also the expansion and pervasive IoT in all fields, the malicious and fraudulent individuals who intended to cheat started to increase as well. Therefore, to protect users (individuals/things) whose security is of prime importance started to protect their information \cite{8,17,18}. So, the two important factors are as follows:
\subsection{Privacy}
Users' private security is indicate the level of accessibility to other users. People can access to others' private information \cite{19}. Some researchers have divided the privacy information into three sections of: infrastructure security, information security and information management \cite{9}. Each user (individual or device) can give access to its sensitive information to others. However, it can limit the access and closes its privacy.
\subsection{Security}
High volume and velocity makes security difficult. This could be investigated from. Users, service rendering centers and computer networks. BD security is quite complicated and still not thoroughly known. Also, technology speed and information volume makes it more difficult. BD security is quite complicated and still not thoroughly known. Also, technology speed and information volume makes it more difficult \cite{8}.\\
Lots of worries exist on information Security and data. As data are stored in voluminous spaces, the accesses of fraudulent people is probable. Giant Companies like Google, Microsoft, YouTube, Skype, etc. Have tried to devise various information Security methods which shows how perilous security protection is in giant servers \cite{18}.\\
Security threats consist of several sections. Dennial of service (DoA) attack is formed in the infrastructure, and attack on encryption function and access control are formed in privacy section \cite{9}. Preventing users against these attacks are security challenges.
\section{Threats and Challenges }
This is the most important part of our paper. Since it deals with the challenge of protection. In this section could be divided into gathering of BD, BD analysis and use of BD. These include: subdivisions as BD challenge, processing challenge and BD management challenge \cite{21}. Please note that: We have pointed out to current challenges. In the section $5$ focuses on how to overcome on challenges related to security and privacy \cite{4,6,8,10,13,20}. We describe some features of BD in the blow:
\begin{itemize}
\item \textbf{Volume:} Volume of data  (terra byte and more) is a great challenge in BD. The diversity of the type of information  environmental information medical info and business information. Facebook, for example produces $500$ terra byte data every day. As it is known from the name BD, the volume counts. We do not set boundaries for the BD, but it is not low-scale data.
\item \textbf{Variety:} The data are different. For example the corresponded data by sensors are many variation (e.g. environment information, sound, image, data, and even noise).
\item \textbf{Velocity:} As mentioned before pointing to diversity and complexity of data structure, if high speed and voluminous data is added, processing becomes a challenging job. Velocity is another feature of BD and its challenge. The processing of high-velocity data is a challenge.
\item \textbf{Variability:} Usually users transmit different data. Google, for instance, receives diverse data from different users and different sources.
in some other resources variability of data is among the four original challenges. We depict a $4V$ in the figure \ref{source}.
%%%%%%%%%%%%%%%%%
\begin{figure*}
\centering
\includegraphics[scale=0.18]{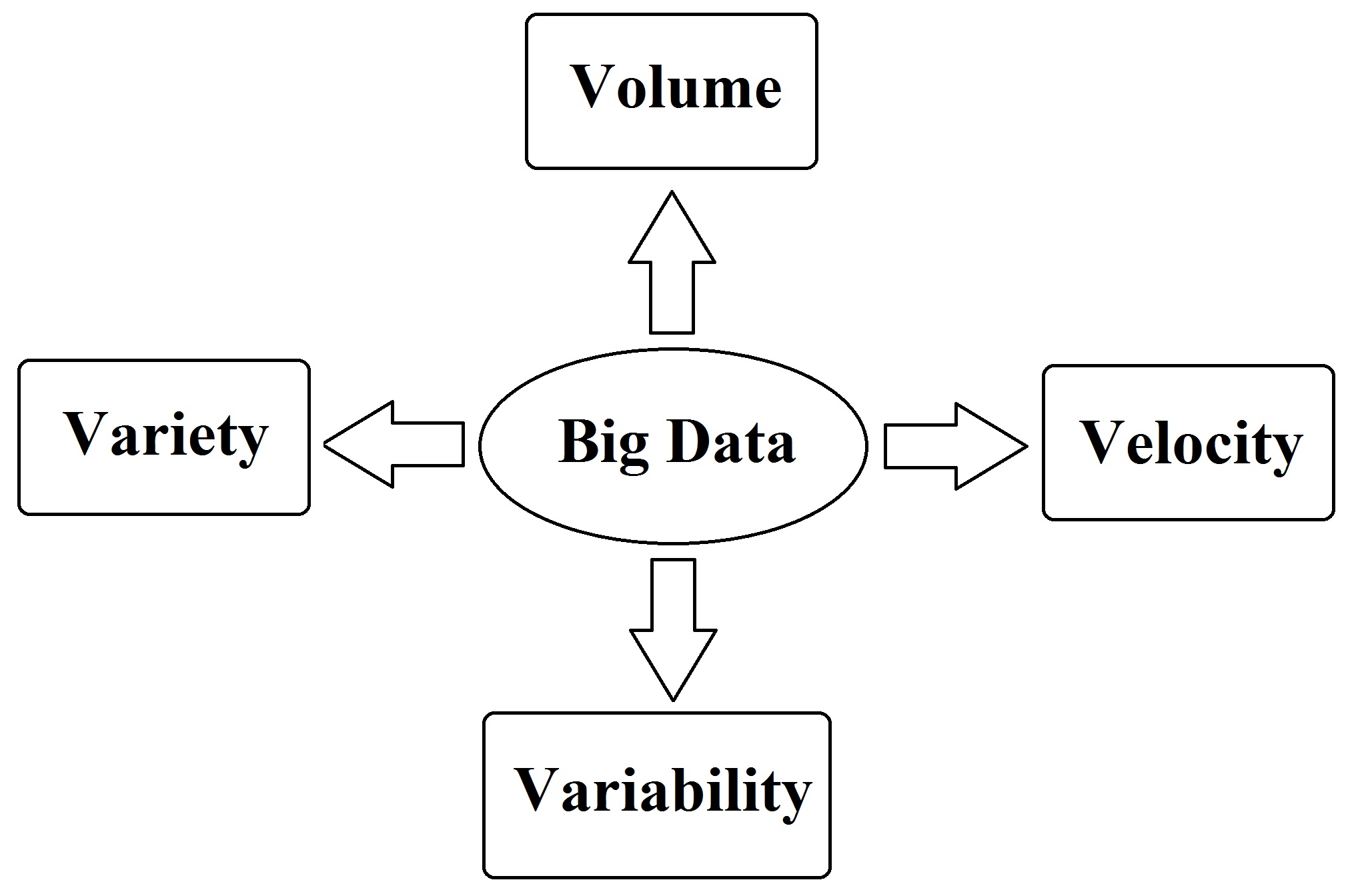}
\caption{The $4$ original challenges of big data (4v) \cite{16}}
\label{bd}
\end{figure*}
%%%%%%%%%%%%%%%%%
\item \textbf{Veracity:} The structure of data is quite complicated and in BD confidence cannot be established.
\item \textbf{Visualization:} Information shall be understandable and eligible. From every sender key information shall be available. For example, eBay has many clients whose info have to be received.
\item \textbf{Value:} The value of data shall be kept and maintained. None of it shall be deleted. It is very important that the value of data does not lost/change. 
\item \textbf{Data acquisition:} This is a challenging job related to information. Gathering from different sources and keep them secure.
\item \textbf{Data Mining and Cleansing:} One of the greatest challenges, is to cleanse the giant pool of info from additional data and choice of correct data required.
\item \textbf{Data aggregation and integration:} This challenge shall be explained by one example: Imagine the giant social media: Twitter. You should answer any tweet by a retweet. Finding the reply to a tweet in this great volume of data. The most important is the answer to the right question which is the duty of twitter server.
\item \textbf{Complexity:} As clear from other explanations, as data volume increases, complexity boosts as well. This could be divided into three parts 1)data complexity, 2)computational complexity, and 3)system complexity.
\item \textbf{Data analysis and modelling:} This challenge points out to some type of sub sets: separation of information, and gathering lost data.
\item \textbf{Data interpretation:} This step looks like image building step important for received data decision.
\item \textbf{Privacy:} As mentioned before, privacy is the most challenging task in digital age. In the section $3$, we discussed the importance of this feature.
\item \textbf{Security:} Keeping users' privacy and preventing from spread of malware. According to the importance of processing of BD, their security should also be considered. It means that the attacker and malicious user have no ability to access the users' sensitive information.
\item \textbf{Data governance:} As BD is increasing, companies and Organizations shall potentially try to manage the data properly, data quality guarantee, data quality improvement and keeping the value of the data is among key elements in information.
\item \textbf{Data and information sharing:} All Organization, Shall coordinate in sharing information.
\item \textbf{Cost:} Data cost are increasing as demands increases. All throughout the world client data shall be supported. For example, Google supports its clients in $13$ centers around the world.
\item \textbf{Data Ownership:} In addition to security and privacy, data ownership is important as well, which shows itself when sharing.
\end{itemize}
\section{Methods to Overcoming on Challenges}
As discussed, BD has a lot of challenges such as 1)security and privacy, 2)data management, and 3)processing challenges.\\
In this paper, we outline the security and privacy implications of some ways to overcome on these challenges in the Internet of Things.  The most basic and primitive goal in security and privacy is to provide all three features of Confidentiality, Antegrity, and Authentication (CIA) \cite{17}.
\paragraph{Using encryption functions} Encountering giant networks like IoT functions and protocols shall be used to comply with all things. Many devices have limited processing power and cannot be encrypted through keys. Many devices have low power. But today many methods are devised for encrypting which is both speedy and secure and low cost \cite{10,17}.
\paragraph{Network security} Protection techniques are vital \cite{10}.
\paragraph{Protecting programs} Program protection is vital, for there are attacks to steal information. There are also many IoT programs which are still nascent in $ID$, digital signature and Each has strength and weaknesses. Sometimes through simple changes the program could be practically useful \cite{10}.
\paragraph{Using secure protocol} To protect  users' privacy and protect them on the network, secure protocols (and lightweight) should be used. There are many protocols for transferring information, user authentication, digital signatures, and so on. Each of them is presented for the purpose of the properties, each of which has strengths and weaknesses that can be secured by analyzing them in detail. Sometimes, with a slight change, the use of a communication protocol can be changed, or by modifying a protocol with a simple technique, it can be used in practice.
\paragraph{Using from lightweight encryption functions} Using cryptographic functions is one of the best ways to protect information. But since the future targets are clear, the largest share of future information belongs to sensors and RFIDs. Therefore, the use of light cryptographic functions is very important. Since most of the elements on the Internet are objects of sensors or smart cards, they can be a symmetric cryptographic key in their production plant.
\paragraph{Smart design of communication protocols} Lightweight protocols and the use of cryptographic functions of the network security network will not always be used, and sometimes it is necessary to use asymmetric encryption, digital signing, and other encryption functions that greatly increase the computing load of the network. Therefore, it is necessary to design a protocol that is the main burden of its calculation on the server servers and users (objects) is powerful. So that poor network elements such as sensors carry less computational burden.
\section{Conclusion}
In this paper, we reviewed and analyzed some of the projects on the IoT with the BD mining approach and we have reviewed the safety and privacy of the users. Eventually, we used several methods, such as the use of encryption functions and protection programs. It is absolutely clear that encryption functions Bring confidentiality. After that, we proposed methods that could provide a safer space for low-power users. The most important of our proposed methods is the smart design of communication protocols, so that the computing and processing load will be borne by service providers (cloud servers). In this method, the light things (e.g. sensors) apply light operation. But, on the other side of protocol, the server uses the same or other cryptographic function. So, the low-power things can run the protocol and be safe.
\section{Future Works}
Given the ever-expanding Internet network of objects in all areas, researchers are thinking of the future of the network. We know that the most important part of the data is large data aggregation, which should be processed after processing and other actions on them. We should know that the most exchanged data  of this network is more than that group in the future. Some scholars will map out the future generations of macroeconomic data into 1) online data networks, 2) cellphone and Internet data objects, 3) geographic information, 4) temporary space data, And 5) flow data and many other data \cite{22}. Methods must be considered for the collection, processing and classification of these categories.\\
One of the most important parts of the today world is industrialize. The industrialization of societies is on the risk, and this trend will continue, and people will be replaced by cars. The vast Internet of objects in the industry can not be ignored, and in the industry, the Internet of Things will play a significant role. The expansion of the industry, the spread of information, as well as the increasing of data sent by the agents in it. Sensors, drives and other components can be referred to  the factors that are constantly sending and receiving data \cite{15}. Securing this volume of equipment and devices for sending and receiving information is very important. Failure to send and receive one of these data may result in a serious problem with production. Therefore, maintaining the security and health of industrial data in the present and future will be very important.\\
 It is anticipated that the number of objects of the Internet service provider will reach a billion by 2030, so companies in this field will be researching. According to HP and Intel, large data management and processing should have three features. First of all, there should be a lot of powerful and high-capacity terminals for data gathering, second, the data produced by the Internet elements of objects is not often  complete, and it should be possible to analyze them properly. Finally, the information gathered from the elements of the Internet of objects is only effective when analyzed \cite{16}.\\
The use of electronic devices is increasing day by day, and as a result, all communities try to make the environment more intelligent and use energy efficiently. Increasing the use of electronic devices (sensors, smart devices and even cloud servers) will require more privacy. In Section 5, we used one of the ways to protect information and privacy using cryptographic functions. An important point is  using its computational load encryption functions. Designers should be careful about the privacy of users (information has been sent and received by objects or individuals) and the optimal use of energy in designing Internet systems and service servers.\\ \\
\textbf{Conflict of Interests}\\ The authors declare that they have no conflict of interests.

\end{document}